\documentclass[preprint,amsmath,amssymb]{revtex4}
\usepackage{graphicx}
\usepackage{dcolumn}
\usepackage{bm}
\begin{document}
\title{A tale of two velocities: Threading vs Slicing}
\author{${\rm R.\;Gharechahi}^{\;(a)}$ \footnote {Electronic
address:~r.gharechahi@ut.ac.ir}, ${\rm M.\;Nouri}$-${\rm Zonoz}^{\;(a)}$\footnote{Electronic
address:~nouri@ut.ac.ir, corresponding author} and ${\rm A. \;Tavanfar}^{\;(b)}$ \footnote{Electronic
address:~alireza.tavanfar@cern.ch}}
\affiliation{(a): Department of Physics, University of Tehran, North Karegar Ave., Tehran 14395-547, Iran.\\ 
(b): International Institute of Physics, Universidade Federal do Rio Grande do Norte, 59078-400 Natal-RN, 
Brazil.}
\begin{abstract}
Two principal definitions of a 3-velocity assigned to a test particle following timelike trajectories in stationary 
spacetimes are introduced and analyzed systematically. These definitions are based on the $1+3$ (threading) and $3+1$ (slicing) 
spacetime decomposition formalisms and defined relative to two different sets of observers.  
After showing that Synge's definition of spatial distance and 3-velocity are equivalent to those defined in the  $1+3$ (threading) formalism,
we exemplify differences between the two definitions, by calculating them for  particles in  circular orbits in axially symmetric 
stationary spacetimes. Illustrating its geometric nature,
the relative linear velocity between the corresponding observers is obtained in terms of the spacetime metric components. Circular particle 
orbits in the Kerr spacetime, as the prototype and the most well known of stationary spacetimes, are examined with respect to these 
definitions to highlight their observer-dependent nature. \\ 
%For orbits close to the marginally bound spherical orbit, the difference between the local velocity measurement 
%and that made by the distant observers could be as high as 20 percent of the velocity of light.
We also examine the Kerr-NUT spacetime in which the NUT parameter contributing to the off-diagonal terms in the metric is mainly interpreted
not as a rotation parameter but as a gravitomagnetic monopole charge. 
Finally, in a specific astrophysical setup which includes rotating black holes, it is shown how these local definitions are related to the velocity measurements 
made by distant observers using spectral line shifts.
\end{abstract}
\maketitle
%----------------------------------------------------------------------------------------
%	INTRODUCTION 
%----------------------------------------------------------------------------------------
\section{Introduction and motivation}
Today the most promising theory of gravity still remains to be Einstein's general theory of relativity which formulates gravity as spacetime geometry. 
After introduction of
any theory, specially the revolutionary ones, an obvious and usually tedious task is to define the observable quantities and their relation to the mathematical 
entities introduced in the formulation 
of that theory. The question of what is measurable and what is not, and indeed what is the meaning of measurement itself, seems to be still one of the puzzling 
issues in quantum mechanics, where one should 
relate laboratory  measurements of phenomena originated at the subatomic world to the concepts and quantities introduced in the quantum mechanical formulation of
the same world. In the case 
of general relativity and its geometrical formulation of gravity, this amounts to finding relation between geometrical entities introduced in the theory and the 
observations made on the large scale world 
or in the presence of strong gravitational fields, where 
this theory is supposed to be at work. Since one is allowed to use different frames and observers attached to them, one should be cautious with another complexity 
which arises naturally due to 
the observer-dependence of the observable quantities.\\
One of the important observable quantities that is dealt with in astrophysics and cosmology, is the 3-velocity of astrophysical objects which is measured either 
through astrometric observations or 
frequency-shifted electromagnetic signals
received from those objects, the so called {\it spectroscopic velocity}. Now one may ask the question what is the relation between these observationally  measured 
values and those defined in the mathematical formulation 
of the theory. \\
Obviously, as long as one  deals with the 3-velocity in the realm of Newtonian 
mechanics or in  special relativity, the usual {\it flat} spacetime definition is employed, the so called {\it coordinate 3-velocity}. But even in this apparently 
simple case, both the mathematical definitions and  
observational methods alike, are not immune from ambiguities arising from different effects entering the measurement process. These should be taken into account to 
give a consistent set of definitions
and measurement procedures \cite{Lind}. When it comes to the curved backgrounds, it is expected that one should deal with a much more difficult task in defining a 
similar concept. 
Obviously if in curved backgrounds we are going to borrow the same basic idea used to define the 3-velocity of a particle 
in Newtonian mechanics, namely spatial distances and time intervals, then one should choose a decomposition scheme to introduce spatial and temporal
intervals in the context of general relativity. \\
The famous example incorporating 3-velocities is the rotation curves of disk galaxies which present strong evidence for the existence of the so called dark matter. 
These are plots of the magnitude of the orbital velocity of visible matter in the galaxies (or clusters of galaxies) versus radial distance from the galactic (cluster) center. Another interesting example includes  
orbital velocities of particles moving in near-circular orbits around rotating black hole candidates, such as in the case of accretion disks around black holes 
produced by the material falling into the black hole 
from its companion star in  black hole-star binaries . 
Although in the first example people mostly use the usual Newtonian definition of a 3-velocity due to the fact that in the galactic scales general relativistic 
effects are negligible 
%\footnote{It should be noted there are studies in which  vacuum and non-vacuum solutions of GR are employed to model galactic dynamics.}
, in the second example one has to employ GR-based calculations 
to account for the strong gravitational field of black holes.
To assign  3-velocities to objects in curved spacetimes by local observers, one first needs to define spatial distances and time intervals between two nearby events 
in the underlying curved spacetime.
Indeed the main idea of any splitting formalism in GR is the introduction of spatial and temporal sections of a  spacetime metric so that one could assign spatial 
distances and time 
intervals to nearby events. There are two well-known decomposition formalisms namely : I- $3+1$ (or slicing formalism) and II- $1+3$ (or threading formalism). In 
the more famous $3+1$ splitting, the spacetime manifold 
is foliated into constant-time hypersurfaces and the spacetime metric is written in terms of the so called lapse function and shift vector \cite{MTW}. On the 
other hand in the $1+3$ formulation 
of spacetime decomposition the propagation of light signals between any two nearby timelike observers
is employed to express the spacetime metric in terms of the so called {\it synchronized proper time} interval and spatial distance \cite{Landau, Zelmanov}. This is 
the same decomposition formalism which is 
employed to  introduce the so called {\it quasi-Maxwell} form of the Einstein field equations in the broader context of {\it gravitoelectromagnetism} \cite{Lynden}. 
In what follows we restrict our attention to stationary 
spacetimes, noting that the two formalisms coincide for static spacetimes. These two different splitting methods lead to two different definitions of 3-velocities 
as measured relative to two different 
sets of observers. Now there are  two questions in order  1- Whether or not and how these two definitions are related? and 2- What are their main differences, 
specially when applied to test particles in 
astrophysically relevant cases of axially symmetric spacetimes such as the Kerr metric?. To simplify things we restrict our study to particles in circular 
orbits in axially symmetric stationary spacetimes, 
noting that most interesting cases including the above mentioned examples fall into this category.
The outline of the paper is as follows. In  sections II  we introduce the $1+3$ spacetime decomposition and its definition of 3-velocity. In section III we will 
briefly discuss Synge's formalism of 
spacetime measurements and show that his measures of spatial and temporal distances, and accordingly that of 3-velocity, are  equivalent to those defined 
in $1+3$ formalism. In section IV we introduce the  
slicing formalisms of spacetime decomposition and its definition of 3-velocity. In section V, after noting that the two definitions of the 3-velocity  are 
defined relative to 
two different sets of observers, we compare them and obtain their relation. These definitions are then calculated in Kerr and Kerr-NUT spacetimes. To relate 
these definitions of 3-velocity to those 
measured by astronomers we discuss a special astrophysical setting in section VI. We summarize and discuss our results in the last section.\\
{\bf Notations}: Following Landau and Lifshitz \cite{Landau} our convention for indices is such that the Latin indices run from 0 to 3 while the Greek ones 
run from 1 to 3. Throughout 
we employ gravitational units in which $c = G =1$ and indices $T$ and $S$ stand for threading and slicing formalisms respectively. 
%%%%%%%%%%%%%%%%%%%%%%%%%%%%%%%%%%%%%%%%%%%%%%%%%%%
\section{Definition of 3-velocity in 1+3 splitting (threading) formalism}
The $1+3$ formulation of spacetime decomposition is the decomposition of spacetime by the so called {\it fundamental observers} in a gravitational field. 
These observers are at fixed spatial 
points in the space defined by the  hypersurfaces $x^0= constant$, so that their timelike world lines, acting as the time lines, 
decompose the underlying spacetime into timelike threads (hence the name threading is justified) \cite{Landau, FN, Mashhoon}. In stationary, asymptotically 
flat spacetimes, these observers 
are at rest with respect to distant observers
in the asymptotically flat region \footnote{The space defined in this way is also called {\it absolutely rigid space} \cite{Tho} or  {\it absolute space} \cite{FN} 
and the 
corresponding observers/frames are alternatively called {\it Lagrange or Killing observers/frames}.}. In this formalism it is the light signal propagation which is 
employed to characterize element of 
spatial distances between two nearby events \cite{Landau}. 
The spacetime metric in this formalism is expressed in the following general form:
\begin{equation}\label{ds0}
d{s^2} = d\tau _{syn}^2 - d{{l_T}^2} = {g_{00}}{(d{x^0} - {g_\alpha }d{x^\alpha })^2} - {{\gamma_T} _{\alpha \beta }}d{x^\alpha }d{x^\beta }
\end{equation}
where ${g_\alpha } =  - \frac{{{g_{0\alpha }}}}{{{g_{00}}}}$ and
\begin{equation}\label{gamma0}
{{\gamma_T} _{\alpha\beta }} =  - {g_{\alpha \beta }} + \frac{{{g_{0\alpha }}{g_{0\beta }}}}{{{g_{00}}}} \;\; ; \;\;  {{\gamma_T} ^{\alpha \beta }} =  - 
{g^{\alpha \beta }}
\end{equation}
%%%%%%%%%%%%%%%%%%%%%%%%%%%%%%%%%%%%%%%%%%
\begin{figure}\label{1}
\includegraphics[scale=0.75]{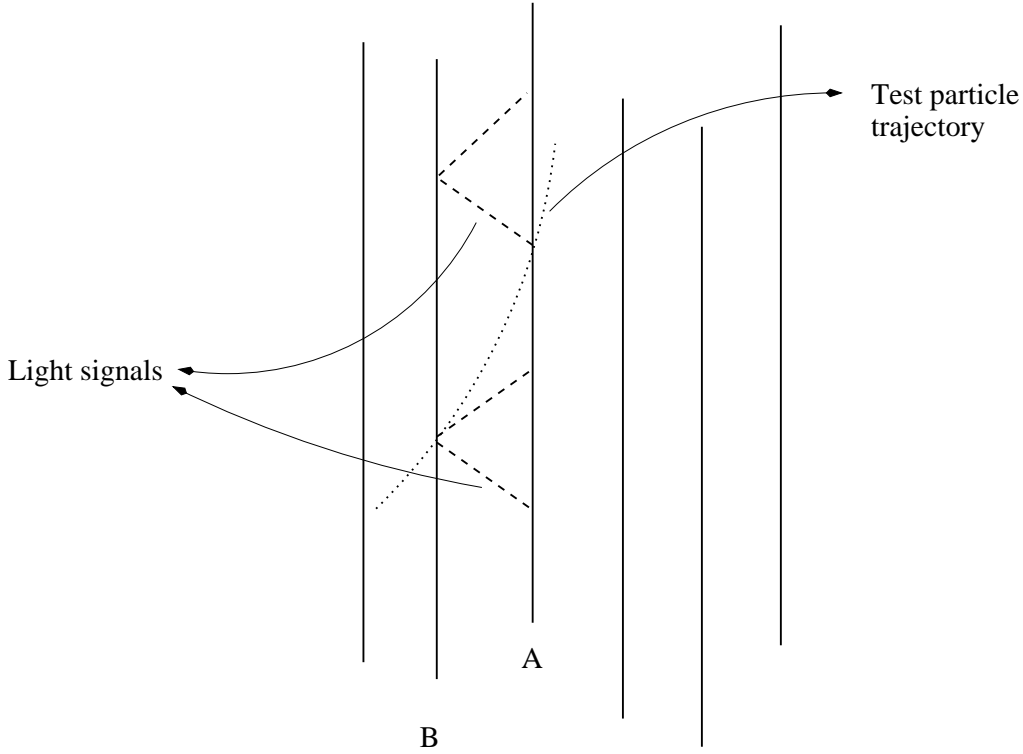}
\caption{A congruence of nearby worldlines of fundamental observers and a test particle crossing two of them (A and B) in the threading formulation of 
spacetime decomposition. Particle 3-velocity 
is defined in terms of the proper time read by clocks synchronized along the particle's worldline by sending and receiving light signals.}
\end{figure}
%%%%%%%%%%%%%%%%%%%%%%%%%%%%%%%%%%%%%%%%%%
is the spatial metric of a 3-space $\Sigma_3$, on which $d{{l_T}}$ gives the element of spatial distance
between any two nearby events. The 3-space $\Sigma_3$ defined in this way is called a {\it quotient space/manifold} and in general is not a submanifold of 
the original 4-d manifold \cite{Nouri,stephani}. \\
Also $d{\tau _{syn}} = \sqrt {{g_{00}}} (d{x^0} - {g_\alpha }d{x^\alpha })$ gives the infinitesimal interval of the so called {\it synchronized proper time} 
between any two events. In other words any two simultaneous events have a world-time difference of $d x^0 = {g}_\alpha dx^\alpha$. In the threading formalism, 
the 3-velocity for a particle is defined
in terms of the synchronized proper time read by clocks synchronized along the particle's trajectory. The origin of this  definition of a time interval could 
be explained through the following procedure. 
If the particle departs from point B (with spatial coordinates $x^\alpha$) at the 
moment of world time ${x^0}$ and arrives at the infinitesimally distant point A (with spatial coordinates $x^\alpha + d x^\alpha$) at the moment ${x^0} + d{x^0}$, 
then to determine the velocity we 
must now take, difference between ${x^0} + d{x^0}$ and the moment ${x^0} - \frac{{{g_{0\alpha }}}}{{{g_{00}}}}d{x^\alpha }$ which is {\it simultaneous} at the point 
B with the moment ${x^0}$ at the 
point A (Fig. 1). This time difference amounts to the synchronized proper time between the two nearby events (departure of the particle from point B and its 
arrival at point A)  and upon dividing 
the infinitesimal spatial coordinate interval $dx^\alpha$ by this time difference the 3-velocity of a particle in the underlying spacetime is given by \cite{Landau}
\begin{equation}\label{velo}
{({v_T})^\alpha } = \frac{{d{x^\alpha }}}{{d{\tau _{syn}}}} = \frac{{cd{x^\alpha }}}{{\sqrt {{g_{00}}} (d{x^0} - {g_\alpha }d{x^\alpha })}}
\end{equation}
Obviously in the case of flat spacetime (i.e when $g_{00}=1 \;\; {\rm and} \;\; g_{0\alpha} = 0$ \footnote{Actually these are conditions for a synchronous 
reference system, 
but the only vacuum stationary spacetime in this system is the flat spacetime \cite {Landau}.}) the above definition 
reduces to the coordinate time velocity defined by $v^\alpha = \frac{dx^\alpha}{dx^0}$.
One of the main features of the $1+3$ formulation is that one could express the Einstein field equations in the so called {\it quasi-Maxwell} form in the
context of {\it gravitoelectromagnetism}. 
Using the 1+3 formalism, it is shown that a test particle moving on the geodesics of a {\it stationary} spacetime, 
depart from the geodesics of the 3-space $\Sigma_3$ as if acted on by the following gravitoelectromagnetic Lorentz-type 3-force \cite{Landau,Lynden},
\begin{equation}
{\bf f}_g = \frac{m_0}{\sqrt{1-{v^2}}}\left( {\bf E}_g + {\bf v}\times \sqrt{g_{00}}{\bf B}_g\right)
\end{equation}
in which the 3-velocity of the particle is defined by \eqref{velo}
and the GE and GM {\it vector fields} are defined as follows,
\begin{gather}
\textbf{B}_g = curl~({\bf A}_g) \\
\textbf{E}_g = -{\bf {\nabla}} \phi.
\end{gather}
Using the definition \eqref{velo} the square of the 3-velocity is then given by,
\begin{equation}
{(v_T)^2} = (\frac{dl_T}{d{\tau _{syn}}})^2 = {{\gamma_T}_{\alpha \beta }}{{v_T}^\alpha }{{v_T}^\beta } = 
\frac{{{{({g_{0\alpha }}d{x^\alpha })}^2} - {g_{00}}{g_{\alpha \beta }}d{x^\alpha }d{x^\beta }}}{{{{({g_{00}}dx^0 + {g_{0\alpha }}d{x^\alpha })}^2}}}.
\end{equation}
For later comparison we note that using \eqref{ds0} and the above relation, the interval $ds$ could be expressed in terms of the velocity in the following form:
\begin{equation}\label{ds00}
d{s^2} = {g_{00}}{(dx^0 + \frac{{{g_{0\alpha }}}}{{{g_{00}}}}d{x^\alpha })^2}[1 - {{v_T}^2}] = d\tau_{syn.}^2 (1-{v_T}^2)
\end{equation}
Restricting our attention to the case of axially symmetric stationary spacetimes in cylindrical coordinates $(t, r, z, \phi)$ with the following general form,
\begin{equation}\label{ax1}
d{s^2} = {g_{tt}}({r,z}){(dt)^2} + 2{g_{t\phi }}({r,z})d{t}d{\phi } + {g_{\alpha \beta }}({r,z})d{x^\alpha }d{x^\beta }  \;\;\;\ ; \;\;\; \alpha,\beta = r, z,\phi 
\end{equation}
we also confine the motion of the particle to a circular orbit around the axis on the  $z=constant$ hypersurface such that the only non-vanishing component of the 
3-velocity has the following form
\begin{equation}\label{m1}
({{v_T}^\phi})^2 = \frac{{g_{t\phi }^2 - {g_{tt}}{g_{\phi \phi }}}}{{{{({g_{tt}} + {g_{t\phi }}\omega)}^2}}}\omega^2.
\end{equation}
where $\omega =\frac{d\phi }{dt}$ is the {\it coordinate} angular velocity of the particle. In the case of flat spacetime (in cylindrical coordinates) it reduces 
to the
expected result ${v_T} ^2 = -g_{\phi \phi}(\frac{{d\phi }}{{dt}})^2 \equiv r^2 \omega^2$ .
%%%%%%%%%%%%%%%%%%%%%%%%%%%%%%%%%%%%%%%%%%%%%%%%%%%%%%%%%%%%%%%%%%%%%%555
\section{A note on Synge's definition of spatial distance and relative velocity}
\begin{figure}\label{2}
\includegraphics[scale=0.5]{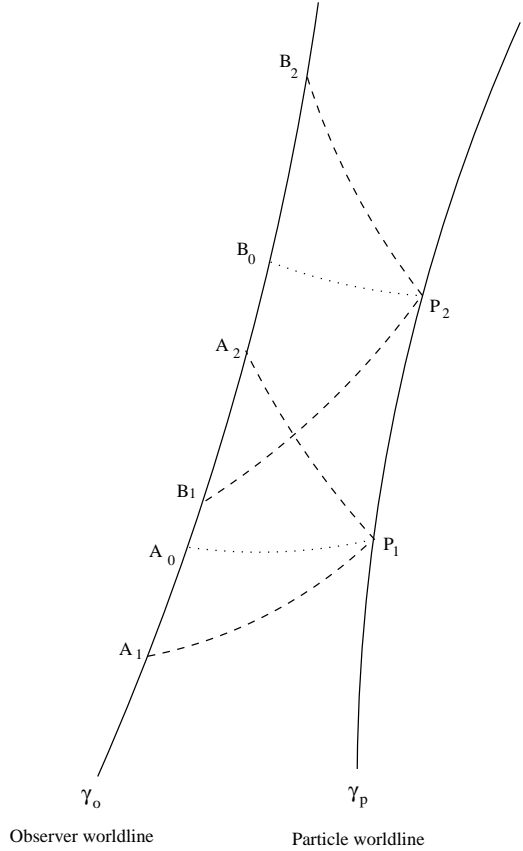}
\caption{The worldlines of an observer and a particle with light signals (dashed lines) sent and received by the observer to the two points ${\rm p}_1$ 
and ${\rm p}_2$ on the particle worldline. 
Also shown are the two points $A_0$ and $B_0$ which are simultaneous with these two points respectively, lying on spacelike curves (dotted lines).}
\end{figure}

Here we digress to discuss Synge's formalism of spacetime measurements and show that it is equivalent to the $1+3$ formalism discussed above. 
Synge \cite{synge} in his very conceptual approach to the space and time measures, 
defines the spatial distance between an observer and a nearby particle as the length of the spacelike geodesic connecting them orthogonally at the position of 
the observer on its timelike worldline. 
The relative (recession) velocity of the particle with respect to the observer is defined by following the particle on its worldline at two successive points  
as the ratio of the difference between the lengths of the 
spacelike geodesics connecting the particle and the observer to the proper time difference for particle positions as measured on the observer's worldline.
De Felice and Clark \cite{defel} generalized Synge's definition of spatial distance, and accordingly his definition of relative velocity, to include the case
of finite 
distance between the observer and the particle where the spacetime curvature is not negligible and should be taken into account. For an observer with a 
timelike geodesic 
worldline $\gamma_o$ \figurename{2} and a particle with worldline $\gamma_p$ at point $p_1$ (the event), they arrive at the following result 
\footnote{To arrive at a simple equation, the authors consider $\gamma_o$ to be 
geodesic itself so that all the first-order
curvature contribution which are coupled to the observer's acceleration vanish. This will not affect our study as we will  be only concerned 
with the case in which the observer and particle are close enough 
to neglet the curvature effects.}, 
\begin{equation}\label{time}
\delta T_{\gamma_o} (A_1, A_2) = 2L (p_1,\gamma_o) - \frac{1}{3} (R_{ijkl}u^i\xi^j u^k \xi^l)L^3(p_1,\gamma_o) + {\cal O}\left(|Riemann|^2\right)
\end{equation}
in which $u^i$ and  $\xi^j$ are the tangents to the observer worldline and the spacelike geodesic joining the observer to the particle respectively, and where
$\delta {T_\gamma}_o (A_1, A_2)$ is the proper time difference on the observer's worldline at points $A_1$ and $A_2$ from which the observer sends and receives 
the light signal to the particle at 
its normal neighborhood.
%%%%%%%%%%%%%%%%%%%%%%%%%%%%%%%%%%%%%%%%%%
%%%%%%%%%%%%%%%%%%%%%%%%%%%%%%%%%%%%%%%%%%
Also $L (p_1,\gamma_o)$ is the spatial distance between the observer and the particle which is taken, according to the Synge prescription, to be the 
length of the spacelike geodesic segment connecting them 
orthogonally at point $A_0$ (i.e $u^i \xi_i |_{A_0} = 0$) on the observer's worldline between the points $A_1$ and $A_2$.
Using the same measure for the spatial distance, particle's radial (recessional) velocity with respect to the observer is given by,
\begin{equation}\label{vel}
V_{rel} = \lim_{p_2 \rightarrow p_1} \frac{\delta L}{\delta_{T_{\gamma_o}}} 
\end{equation}
Obviously in the limit where the two worldlines are so close that the spacetime curvature could be neglected one arrives at,
\begin{equation}\label{time1}
L (p_1,\gamma_o) \approx \frac{\delta T_{\gamma_o} (A_1, A_2)}{2}.
\end{equation}
At the same limit, as shown by de Felice and Clarck (also implicitly by Synge himself), the spatial distance and time interval between the observer and 
the particle at generic events $A$ and $p$, as measured by 
the observer, reduce to the following expressions,
\begin{equation}\label{time2}
(\delta L_{u}(A,p))^2 = h_{ij} dx^i dx^j + {\cal O}(dx^4) \;\;\;\;\;  ;  \;\;\;\;\; (\delta T_{u} (A_0, A_1))^2 = u_i u_j dx^i dx^j |_A + {\cal O}(dx^4)
\end{equation}
in which $h_{ij} = g_{ij} - u_i u_j $ is the so called {\it projection tensor}. One can easily show that in the case of stationary spacetimes and in a coordinate 
system adapted to the timelike Killing vector, in which 
$u_i \doteq \sqrt{g_{00}}(1 , \frac{g_{0\alpha}}{g_{00}})$ is the 4-velocity of the {\it Killing observers}, the above equations reduce to those defined in the 
$1+3$ decomposition 
formalism (also called projection formalism \cite{stephani}), namely equations \eqref{ds0} and \eqref{gamma0} \cite{Nouri}.
This could also be seen from the fact that according to the above setting, the points $A_0$ and $B_0$ (being on the joining spacelike geodesic segments) 
are the events on the observer worldline which are simultaneous 
with the events $p_1$ and $p_2$ on the particle's world line. This concept of simultaneity was the same used to define the synchronized proper time in 
the $1+3$ decomposition 
formalism. Schematically, neglecting the spacetime curvature for nearby timelike observer/particles allows one to use locally
straight lines to represent the observer worldlines as well as 45 degree lines representing the light signals in \figurename{1} \cite{Landau}. 
%%%%%%%%%%%%%%%%%%%%%%%%%%%%%%%%%%%%%%%%%%%%%%%%%%%
\section{Definition of 3-velocity in the 3+1 splitting (slicing) formalism}
\begin{figure}\label{3}
\includegraphics[scale=0.75]{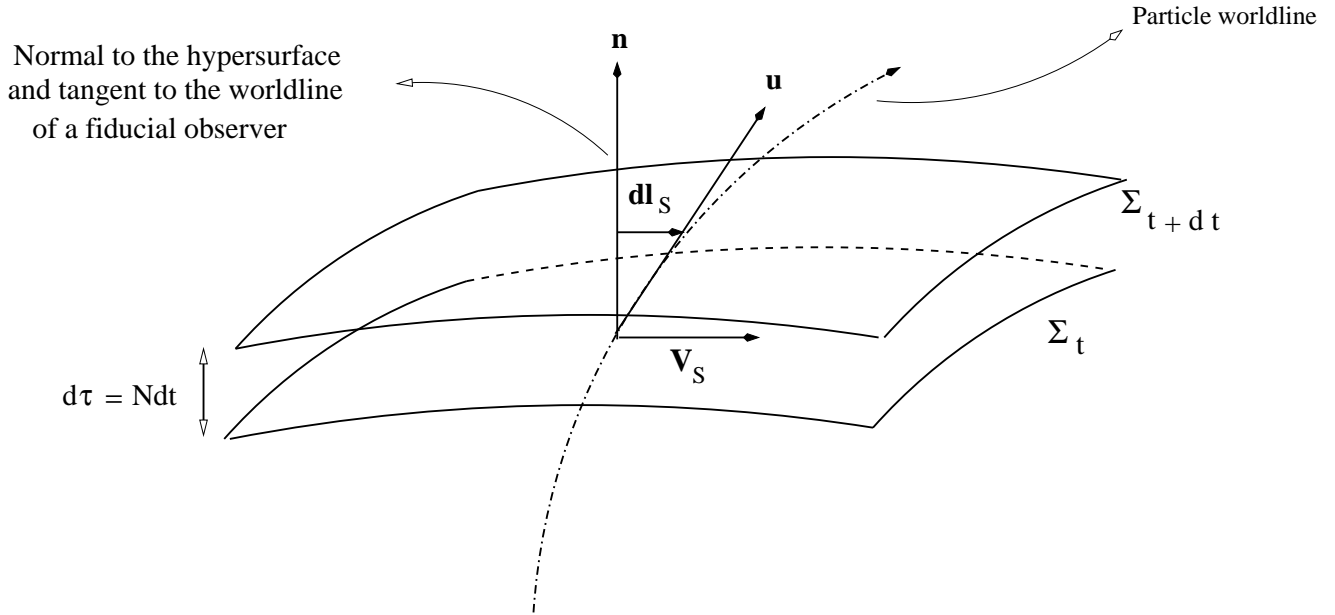}
\caption{Spacetime decomposition in the slicing formalism in terms of the lapse function and shift vector. Proper time and the differential element of
spatial distance used in the definition of the 
3-velocity are also shown.}
\end{figure}

In the $3+1$ splitting, the spacetime manifold is foliated into constant-time hypersurfaces ${\Sigma _t}$
and the (stationary) spacetime metric is written in terms of the so called lapse function $N(x^\alpha)$ and shift vector $N^\beta(x^\alpha)$ in the 
following form \cite{MTW};
\begin{equation}
d{s^2} = {g_{ab }}d{x^a }d{x^b } =  ({N ^2} - {N ^\alpha}{N _\alpha})d{{x^0}^2} - 2{N _\alpha} dx^0 d{x^\alpha} - {{\gamma_{S}} _{\alpha \beta}}d{x^\alpha}d{x^\beta}
\end{equation}
in which
\begin{equation}\label{gamma1}
{{\gamma_{S}}_{\alpha\beta}} =  - {g_{\alpha \beta }} \;\; ; \;\;  {{\gamma_S} ^{\alpha \beta }} =  - {g^{\alpha \beta }} + 
\frac{{{g^{0\alpha }}{g^{0\beta }}}}{{{g^{00}}}}
\end{equation}
are the $3+1$ counterparts of \eqref{gamma0} and
\begin{equation}\label{LSH}
N_\alpha = -g_{0\alpha}\;\;\; ; \;\;\; N^2 =  \frac{1}{g^{00}}
\end{equation}
are the shift vector and lapse function respectively. It is also noted that the lapse function $N$ measures proper time between two neighboring spacelike 
hypersurfaces (\figurename{3})
$d{\tau ^2} =  {N ^2}({x^\alpha})d{{x^0}^2}$ while the shift vector relates spatial coordinates between them  
\[dx^\alpha = x_{x^0 + \delta x^0}^\alpha - x_{x^0}^\alpha =  - {N ^\alpha}({x^\alpha}) dx^0\].
Now the metric could also be written in the following decomposed form 
\begin{equation}\label{ds1}
d{s^2} = d\tau_S^2 - d{{l_S}^2} = {N ^2} d{{x^0}^2} - {{\gamma_{S}} _{\alpha \beta}}(d{x^\alpha + N^\alpha dx^0})(d{x^\beta}+ N^\beta dx^0)
\end{equation}
showing that spatial distances over the spacelike hypersurfaces ${\Sigma _t}$  are given by
\begin{equation}
d{{l_S}^2} = {{\gamma_{S}} _{\alpha \beta}}(d{x^\alpha + N^\alpha dx^0})(d{x^\beta}+ N^\beta dx^0).
\end{equation}
Also it should be noted that the indices of 3-vectors such as the shift vector ${N^\alpha}$ are raised and lowered by the same spatial 
metric ${{\gamma_{S}} _{\alpha \beta}}$.\\
In this formalism for a particle with 4-velocity $u^a$, its 3-velocity $ \upsilon^\mu$  is defined as the projection of the particle's 4-velocity 
onto the above mentioned spatial slices with respect to a time coordinate reference. To this end the timelike and future-directed unit vectors normal 
to the slicing hypersurfaces (denoted by $n^a$) 
are taken as the 4-velocity of the so called {\it Eulerian observers} \footnote{These observers are called Eulerian as they are at rest on {\it spatial slices} 
in the same way that in fluid 
mechanics the Eulerian observers are at rest in {\it space} and do not move with the fluid. To lift any ambiguities on  the analogy made, it should be noted
that in the case of axially symmetric
stationary spacetimes these observers are co-moving with the {\it spacetime fluid}, i.e dragged by it. In other words the analogy is about the observers and 
their relation to the underlying 3-space 
and does not extend to the two fluids.}, relative to whom the 3-velocity is defined as follows
\begin{equation}
{v_S}^\alpha = \frac{d{l_S}^\alpha}{{d\tau }}
\end{equation}
in which the norm of the displacement vector $d{\bf l}_S$ measures the distance between two spatial positions of the particle on the two  hypersurfaces 
${\Sigma _t}$ and ${\Sigma _{t+dt}}$ relative 
to the Eulerian observer. In the same way 
$d\tau $ measures the proper time on the Eulerian observer's worldline corresponding to the  proper time measured on the particles worldline crossing the 
same hypersurfaces \footnote{By the definitions 
of Eulerian observers and the lapse function it is obvious that the same quantity also accounts for the proper time between two hypersurface, i.e $d\tau = Ndt$.}. 
In this way $v$ is tangent to the hypersurface ${\Sigma _t}$ (\figurename{3}). In other words,  
from the Eulerian observers' point of view, the hypersurfaces ${\Sigma _t}$, are the set of all simultaneous events \cite{Tho}.
%%%%%%%%%%%%%%%%%%%%%%%%%%%%%%%%%%%%%%%%%%%%%
%%%%%%%%%%%%%%%%%%%%%%%%%%%%%%%%%%%%%%%%%%%%%
One can show that the 3-velocity, in terms of the coordinate-time velocity, could be written in the following form \cite{Tho,Gou}
\begin{equation}\label{v1}
{v_S}^\alpha = \frac{{1}}{N}(\frac{dx^\alpha}{dt} + N^\alpha)
\end{equation}
and its square, in terms of the four dimensional metric components, is given by
\begin{equation}\label{v11}
{(v_S)^2} =  {{\gamma_S} _{\alpha \beta }}{{v_S}^\alpha }{{v_S}^\beta } = g_{\alpha\beta}\frac{dx^\beta}{dx^0} \left(2g^{0\alpha} -
g^{00}\frac{dx^\alpha}{dx^0}\right) + g_{0\beta}g^{0\beta}       
\end{equation}
in which we used \eqref{gamma1}. Indeed using \eqref{v1} one can express the metric \eqref{ds1} in the following equivalent form 
\begin{equation}\label{ds2}
d{s^2} = {N ^2} d{{x^0}^2} - {dl_S}^2 = d\tau^2 (1- v_S^2)
\end{equation}
which is obviously the $3+1$ counterpart of \eqref{ds00}.
Again restricting our attention to the case of axially symmetric stationary spacetimes in cylindrical coordinates, we  can write
the line element in its general $3+1$  decomposed form \eqref{ds1} as follows,
\begin{equation}\label{ds21}
d{s^2} = \left(g_{tt} - \frac{ g^2_{t\phi}}{g_{\phi\phi}}\right) dt^2 + g_{\phi\phi}\left(d\phi+\frac{g_{t\phi}}{g_{\phi\phi}}dt\right)^2 + g_{rr} dr^2 + g_{zz} dz^2.
\end{equation}
Restricting the motion of the particle to a circular orbit at $z=constant$ hypersurface and reading the lapse function  and shift vector from the above 
form, the squared 3-velocity in this case is given by
\begin{equation}\label{m3} 
({v_S})_\phi ^2 = \frac{{g_{\phi \phi }^2}}{{g_{t\phi }^2 - {g_{tt}}{g_{\phi \phi }}}}{(\omega + \frac{{{g_{t\phi }}}}{{{g_{\phi \phi }}}})^2}
\end{equation}
where, as in the previous section, $\omega = \frac{d\phi}{dt}$ is the only nonzero component of the {\it coordinate} 3-velocity. It is noted ${v_S} =0$ 
for $\omega = - \frac{{{g_{t\phi }}}}{{{g_{\phi \phi }}}}$
allowing the natural interpretation that $- \frac{{{g_{t\phi }}}}{{{g_{\phi \phi }}}}$ is the (dragging) angular velocity induced by the spacetime geometry 
\cite{Landau}.\\
Now having these two different values for the norm of the 3-velocity of a particle in circular motion in stationary, axially symmetric spacetimes we can 
compare them in different
backgrounds. This will be done in the next section.
%%%%%%%%%%%%%%%%%%%%%%%%%%%%%%%%%%%%%%%%%%%%%%%%%%%%%%%%%%%%%%%%%%%%%%%%%%%%%%%
\section{Comparing the two approaches in axially symmetric stationary spacetimes}
To sum up the results in the last three sections, one can argue that in the threading formalism it is the {\it fundamental observers} who choose/define 
what should be 
the {\it local } space over which physical quantities are projected and measured, 
whereas in the slicing formalism it is the slicing  
subspaces (spacelike hypersurfaces) which choose/define the corresponding {\it Eulerian observers}, also called {\it fiducial observers} for the obvious 
reason \footnote{In the special case when 
the underlying spacetime is stationary and axially symmetric
the local frames carried by these observers are also called {\it Locally Non-Rotating Frames} (LNRF).}. In other words in the threading formalism fundamental 
objects are the observers and then, 
with respect to them one defines {\it the
space}.  But in the slicing formalism one first defines what should be the space and then according to those spatial sections defines the (fiducial) observers. 
This way of explaining the difference 
between the two formalisms actually highlights
the important role of the observers in both the definition and measurement of the spatio-temporal quantities such as the 3-velocity of a particle in a 
gravitational field discussed in the previous sections. 
Indeed this is mathematically reflected in the following two metric forms in the two formalisms
\begin{equation}\label{dsn1}
d{s_T^2} =  {g_{00}}{dx_T^0}^2 - {{\gamma_T} _{\alpha \beta }}d{x^\alpha }d{x^\beta } = {g_{00}}{(d{x^0} + \frac{g_{0\alpha}}{g_{00}} d{x^\alpha })^2} + 
({g_{\alpha \beta }} - 
\frac{{{g_{0\alpha }}{g_{0\beta }}}}{{{g_{00}}}})dx^\alpha dx^\beta
\end{equation}
and
\begin{equation}\label{dsn2}
{ds_S^2} = {g_{00}}_S {dx^0}^2 - {\gamma_{S}} _{\alpha \beta}dx_S^\alpha dx_S^\beta  = (g_{00} + {N ^\alpha}{N _\alpha})d{{x^0}^2} +  
{g _{\alpha \beta}}(dx^\alpha + N^\alpha dx^0)(dx^\beta + N^\beta dx^0).
\end{equation}
In the $1+3$ form the cross terms are absorbed to define a new coordinate time difference $dx_T^0 = d{x^0} + \frac{g_{0\alpha}}{g_{00}} d{x^\alpha }$ 
and the corresponding  element of a proper time
(i.e {\it the synchronized proper time} $d\tau_{sync}$), by using the old time-time component of the metric( i.e $g_{00}$). On the other hand the same 
terms are borrowed to define a  
spatial metric ${\gamma_T}_{\alpha\beta}$ and spatial distance using the old spatial coordinate differences, namely $dx^\alpha$.\\
In the $3+1$ form the cross terms are absorbed to define a new time-time component of the metric ${g_{00}}_S = g_{00} + {N ^\alpha}{N _\alpha}$ and the 
corresponding element of proper time ($d\tau_S$), 
by using the old coordinate time difference, namely $dx^0$. On the other hand the same 
terms are also borrowed to define a new spatial coordinate difference $dx^\alpha_S = dx^\alpha + \frac{g^{0\alpha}}{g^{00}}dx^0$ and the spatial distance 
using the old space-space components of the 
metric, namely $g_{\alpha\beta}$.\\
By the same token it is obvious that in the static spacetimes (i.e when $g_{0\alpha} = 0$) the two sets of observers coincide and both formalisms arrive 
at the same result for a particle velocity, 
which in the case of axially symmetric, stationary spacetimes reduce to 
\begin{equation}
(v_S)^2 = (v_T)^2 = - \frac{{{g_{\phi \phi }}}}{{{g_{tt }}}} {\omega^2}.
\end{equation}
The fact that the two definitions introduced are relative to two different sets of observers and consequently should lead to different velocity measures
is more highlighted when one considers the case of a particle with $\omega \equiv \frac{d\phi}{dt}= 0$, i.e a particle with zero 
coordinate velocity. In this case it is noted that equations \eqref{m1} and \eqref{m3} reduce to 
\begin{equation}\label{zero}
v_T = 0 \;\;\;\; ; \;\;\;\; ({v_S})_\phi ^2 = \frac{{g_{t \phi }^2}}{{g_{t\phi }^2 - {g_{tt}}{g_{\phi \phi }}}}
\end{equation}
the first of which is expected intuitively since relative to a fundamental observer, by its definition,  another fundamental observer's velocity (treated as a 
test particle with zero {\it coordinate velocity}) is zero. \\
On the other hand, as noted before,  if we take $\omega = - \frac{g_{t\phi}}{g_{\phi \phi }}$ i.e for a particle moving with the spacetime fluid (dragged by the 
geometry) then 
\begin{equation}\label{zero2}
v_S = 0 \;\;\;\; ; \;\;\;\; ({v_T})_\phi ^2 = \frac{{g_{t \phi }^2}}{{g_{t\phi }^2 - {g_{tt}}{g_{\phi \phi }}}}.
\end{equation}
Again the first result is expected intuitively since  a particle moving with a coordinate velocity equal to the spacetime
angular velocity should be at rest relative to an (Eulerian) observer comoving with the spacetime fluid having the same angular velocity. In other words  an observer 
attached to this particle is an Eulerian observer. \\
The above result is an interesting one showing that the fundamental observers and Eulerian observers which move relative to each other with the 
 {\it angular velocity}  $\omega = - \frac{g_{t\phi}}{g_{\phi \phi }}$ have the relative {\it linear velocity} whose magnitude is given by, 
\begin{equation}\label{vrel}
({u_{TS}})_\phi = \frac{|{g_{t \phi }|}}{\sqrt{{g_{t\phi }^2 - {g_{tt}}{g_{\phi \phi }}}}}.
\end{equation}
Since both the fundamental and Eulerian observers are local, the above assertion could be proved simply by applying the relativistic velocity addition in the 
following form, 
\begin{equation}\label{vadd}
v_T = \frac{u_{TS}+v_S}{1 + u_{TS} v_S},
\end{equation}
in which $v_T$ and $v_S$ are the threading and slicing 3-velocities. It is an easy task to check that Equations \eqref{m1} and \eqref{m3} satisfy the above relation 
with  $u_{TS} = - \frac{|{g_{t \phi }|}}{\sqrt{{g_{t\phi }^2 - {g_{tt}}{g_{\phi \phi }}}}}$ \footnote{The minus sign could be traced back to the fact that for an  
Eulerian observer corotating with the test particle, the fundamental observer moves in the opposite sense.} as the relative linear velocity between the two
observers (frames).
The above results and their interpretations become more clear when we consider explicit examples in the following subsections. 
%%%%%%%%%%%%%%%%%%%%%%%%%%%%%%%%%%%%%%%%%%%%%%%%%%%5
\subsection{Kerr spacetime}
Rotating stars and rotating black holes are among the most important objects in the forefront of the  astrophysical observations and so any observable 
quantity related  directly and indirectly  to the rotation parameter of these objects is also of utmost importance. The spacetime metric around such sources 
is modeled by the Kerr metric 
which represents the axially symmetric, stationary spacetime around a rotating source. It has the following form in the Boyer-Lindquist coordinates,
\begin{equation}
d{s^2} = (1 - \frac{{2Mr}}{{{\rho ^2}}})d{t^2} + \frac{{4Mar{{\sin }^2}\theta }}{{{\rho ^2}}}dtd\phi  - \frac{{{\rho ^2}}}{\Delta }d{r^2} - {\rho ^2}d{\theta ^2} - ({r^2} + {a^2} + 
\frac{{2M{a^2}r{{\sin }^2}\theta }}{{{\rho ^2}}}){\sin ^2}\theta d{\phi ^2}
\end{equation}
where
\begin{equation}
{\rho ^2} = {r^2} + {a^2}{{\cos }^2}\theta \;\;\;, \;\;\; \Delta  = {r^2} - 2Mr + {a^2}
\end{equation}
Here we restrict our attention to the orbits of particles in the equatorial plane ($\theta  = \frac{\pi }{2}$) for which the spacetime metric is given by
\begin{equation}
d{s^2}(\theta = \frac{\pi}{2}) = (1 - \frac{{2M}}{r})d{t^2} + \frac{{4Ma}}{r}dtd\phi  - ({r^2} + {a^2} + \frac{{2M{a^2}}}{r})d{\phi ^2} - 
\frac{{{r^2}}}{\Delta }d{r^2}.
\end{equation}
In this special case the timelike and null orbits in the Kerr geometry are described by two constants of motion which correspond to the total 
energy and angular momentum along the symmetry axis \cite{Carter,Bard}.
It is also noted that in this case we are allowed to use  
formulas obtained in previous sections in cylindrical coordinate due to the fact that the Boyer-Lindquist coordinates coincide with cylindrical 
coordinates in the equatorial plane.
We further restrict our attention to the circular orbits which observationally constitute an important class of orbits in this spacetime specially 
in the study of accretion disks around rotating black holes.\\
In the case of Kerr metric the two sets of fundamental observers and Eulerian observers are called {\it Zero Angular Velocity Observers} (or ZAVOs) 
and {\it Zero Angular Momentum Observers} (or ZAMOs) respectively. 
As shown previously these two sets of observers and their frames move relative to each other with the  angular velocity $ - g_{t \phi} / g_{\phi \phi}$.
In this case for test particles on circular orbits (as timelike geodesics), the only non-vanishing component of coordinate (angular) velocity  
is given by \cite{Bard}
\begin{equation}
\Omega  \equiv \frac{{d\phi }}{{dt}} = \mp\frac{{{M^{1/2}}}}{{{r^{3/2}} \mp a{M^{1/2}}}},
\end{equation}
where the upper sign refers to retrograde (counter-rotating) orbits while the lower one refers to the direct (co-rotating) orbits. 
This quantity is also called Keplerian angular velocity \cite{Abra} for the obvious reason that for $a=0$ it reduces to $\Omega= \frac {M^{1/2}}{r^{3/2}}$ 
which is the Keplerian angular velocity for a test particle in a circular orbit around
a central Mass $M$ in Newtonian gravity. Circular orbits exist in the region $r > {r_{ph}}$, with $r_{ph}$ denoting the radial coordinate of the innermost boundary 
of the timelike circular orbits, the so called {\it circular photon orbit} \cite{Bard},
\begin{equation}
{r_{ph}} = 2M\left\{ {1 + \cos \left[ {\frac{2}{3}arc\cos \left( {\pm\frac{a}{M}} \right)} \right]} \right\}.
\end{equation}
In the  case of circular orbits in the equatorial plane, the only non-vanishing component of the 3-velocity is given by ${v}^\phi$. Now
in the $3+1$ splitting formalism the 3-velocity for direct orbits (into which we will restrict our consideration in what follows), measured relative to the 
ZAMO  has the following form
\begin{equation}
{v_S}^\phi  = \frac{1}{{r\sqrt \Delta  }}\left( {\left( {{r^3} + {a^2}r + 2M{a^2}} \right)\Omega - 2Ma} \right).
\end{equation}
Substituting $\Omega$ into the above equation we end up with
\begin{equation}\label{vf}
{v_S}^\phi  = \frac{{{M^{1/2}}({r^2} - 2a{M^{1/2}}{r^{1/2}} + {a^{2}})}}{{\sqrt {{r^2} - 2Mr + {a^2}} ({r^{3/2}} + a{M^{1/2}})}}
\end{equation}
In the $1+3$ splitting formalism the 3-velocity (in direct orbits) measured relative to the fundamental observers  at rest in the Kerr geometry has the form
\begin{equation}
{v_T}^\phi  = \frac{{{{\left( {{r^2} - 2Mr + {a^2}} \right)}^{1/2}}}}{{\left( {1 - \frac{{2M}}{r} + \frac{{2Ma}}{r}\Omega } \right)}}\Omega
\end{equation}
again with substituting the Keplerian angular velocity as the coordinate angular velocity, we have
\begin{equation}\label{vt}
{v_T}^\phi  = \frac{{{{(M{r^2} - 2{M^2}r + {a^2}M)}^{1/2}}}}{{({r^{3/2}} + a{M^{1/2}})(1 - \frac{{2M}}{r} + \frac{{2{M^{3/2}}{a}}}{{{r^{5/2}} + ar{M^{1/2}}}})}}
\end{equation}
Comparing equations equations \eqref{vf} and \eqref{vt}, the following points are noteworthy:\\
I-As expected, for $r \gg 1 $ (or $\frac{M}{r}, \frac{a}{r} \ll 1$) , in both formalisms the 3-velocities reduce to the  Newtonian value 
$v^\phi = r \frac{{{M^{1/2}}}}{{{r^{3/2}} + a{M^{1/2}}}} \equiv r\Omega$. \\
II-For $a=0$ both reduce to the same 3-velocity  $v^\phi = \frac{M^{1/2}}{(r-2m)^{1/2}}$ corresponding to the Schwarzschild case.\\
III-By equation \eqref{vrel} the two observers move relative to each other with the following linear velocity \cite{FN}
\begin{equation}
u_{TS} = \frac{2Mra \sin \theta}{\rho^2\Delta^{1/2}}|_{(\theta = \pi/2)} = \frac{2Ma }{r \Delta^{1/2}}.
\end{equation}
%%%%%%%%%%%%%%%%%%%%%%%%%%%%%%%%%%%
\begin{figure}\label{4}
\includegraphics[scale=1.2]{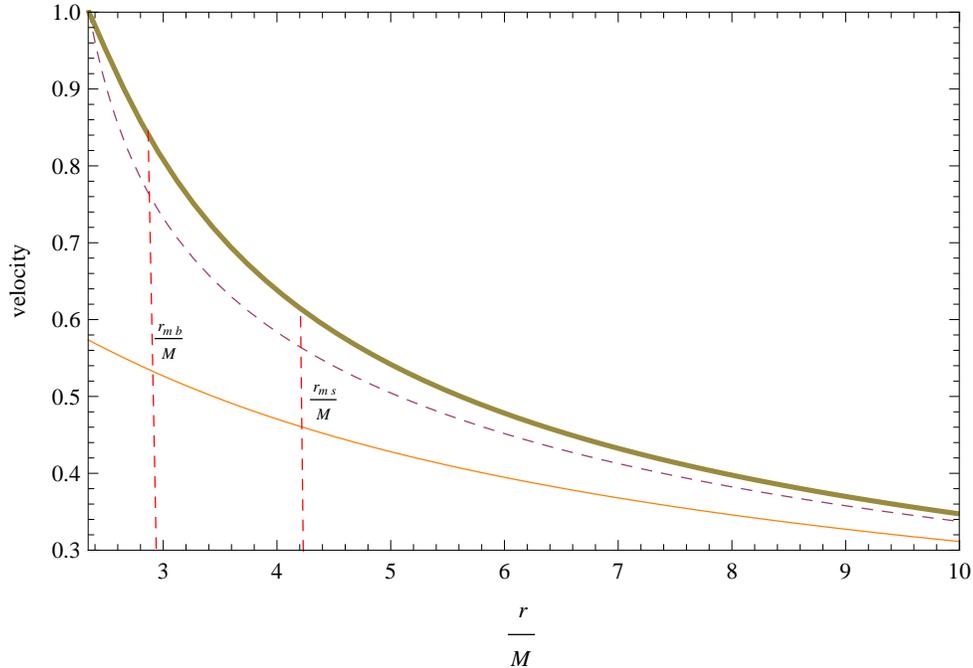}
\caption{Slicing 3-velocity (dashed line), threading 3-velocity (thick solid line) and coordinate 3-velocity (thin solid line) of particles in circular motion 
in the equatorial plane of a Kerr (black) hole with $a=M/2$. Also shown are the  
velocities for the normalized radial coordinates corresponding to the marginally bound ($r_{mb}$) and marginally stable ($r_{ms}$) circular orbits.}
\end{figure}
%%%%%%%%%%%%%%%%%%%%%%%%%%%%%%%%%%%%
In \figurename{4}, the above two 3-velocities along with the coordinate velocity $r\Omega$ (also could be called Keplerian orbital velocity) are depicted as functions 
of the normalized radial coordinate ($\frac{r}{M}$). In both formalisms the 3-velocity of particles in circular orbits 
approach the velocity of light on the photon circular orbit $(r_{ph})$ and tend to zero at infinity.
%%%%%%%%%%%%%%%%%%%%%%%%%%%%%%%%%5
\subsection{Kerr-NUT spacetime}
\begin{figure}\label{5}
\includegraphics[scale=1]{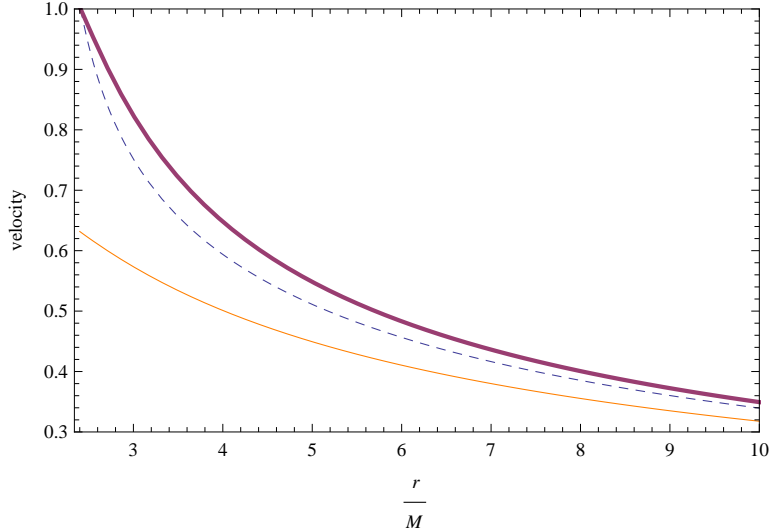}
\caption{Slicing 3-velocity (dashed line), threading 3-velocity (thick solid line) and coordinate 3-velocity (thin solid line) of particles in circular motion 
in the equatorial plane 
of a Kerr-NUT spacetime with $a=M/2$ and $n=M/4$.}
\end{figure}

Kerr-NUT spacetime is a stationary, axially symmetric solution of Einstein vacuum field equations with three parameters $M$, $a$ and $n$ corresponding to mass, 
rotation and the NUT charge (also called magnetic mass) respectively. The last two
parameters source the cross terms in the spacetime metric and it is naturally expected that they also contribute into the difference between the definitions of the two velocities 
discussed in the previous examples. This spacetime in the Schwarzschild-type coordinates has the following form
\begin{equation}\label{kerr}
d{s^2} =  - \frac{\Delta }{{{p^2}}}{(dt - Ad\phi )^2} + \frac{{{p^2}}}{\Delta }d{r^2} + {p^2}d{\theta ^2} + \frac{1}{{{p^2}}}{\sin ^2}\theta {(adt - Bd\phi )^2}
\end{equation}
where
\[\begin{array}{*{20}{c}}
{\Delta  = {r^2} - 2Mr + {a^2} - {n^2},}&{{p^2} = {r^2} + {{(n + a\cos \theta )}^2}}\\
{B = {r^2} + {a^2} + {n^2},}&{A = a{{\sin }^2}\theta  - 2n\cos \theta }
\end{array}\]
In the equatorial plane $\theta  = \frac{\pi }{2}$, the line element has the form
\begin{equation}
d{s^2} =  - \frac{{{r^2} - 2Mr - {n^2}}}{{{r^2} + {n^2}}}d{t^2} - \frac{{4a(Mr + {n^2})}}{{{r^2} + {n^2}}}d\phi dt + \frac{{{B^2} - {a^2}\Delta }}{{{r^2} + 
{n^2}}}d{\phi ^2} + \frac{{{r^2} + {n^2}}}{\Delta }d{r^2}.
\end{equation}
Circular orbits in the equatorial plane in Kerr-NUT spacetime have been studied in \cite{Cha}. The angular velocity of a test particle  moving on the 
circular orbits around the symmetry axis 
and is observed by stationary observers at infinity is given by
\begin{equation}
\Omega  = \frac{{\sqrt {{u^3}\left[ {M\left( {1 - {n^2}{u^2}} \right) + 2{n^2}u} \right]} }}{{1 + {n^2}{u^2} \mp a\sqrt {{u^3}\left[ {M\left( {1 - {n^2}{u^2}} \right) 
+ 2{n^2}u} \right]} }}
\end{equation}
where $u = \frac{1}{r}$.
This formula for $ n=0 $ and $ a=0$ gives Keplerian angular velocities in the Kerr and NUT spacetimes respectively. These circular orbits exist only for the region 
$ r>{r_{ph}}$, 
where $ {r_{ph}} $ is again the radius of the photon circular orbit but now given as the solution to the following equation \cite{Prad}
\begin{equation}
r^3 - 3Mr^2 - 3n^2 r \pm 2a \sqrt{r(Mr^2 + 2n^2r - Mn^2)} + Mn^2 = 0.
\end{equation}
Substituting the components of metric \eqref{kerr} in the 3-velocity definitions in $1+3$ and $3+1$ splitting formalisms and employing the above formula for 
the angular coordinate velocity we 
have the following threading and slicing 3-velocities in the Kerr-NUT spacetime,
\begin{equation}
v_S^\phi  = \frac{{ \mp \sqrt {u\left[ {M\left( {1 - {n^2}{u^2}} \right) + 2{n^2}u} \right]} \left[ {1 + ({a^2} + {n^2}){u^2}} \right] - 
2a{u^2}(M + {n^2}u)}}{{\sqrt {1 - 2Mu + 
\left( {{a^2} - {n^2}} \right){u^2}} \left[ {1 + {n^2}{u^2} \mp a\sqrt {{u^3}\left[ {M\left( {1 - {n^2}{u^2}} \right) + 2{n^2}u} \right]} } \right]}}
\end{equation}

\begin{equation}
v_T^\phi  = \frac{{\left( {{r^2} + {n^2}} \right)\sqrt {{r^2} - 2Mr + {a^2} - {n^2}} \left( {\frac{{\sqrt {{u^3}
\left[ {M\left( {1 - {n^2}{u^2}} \right) + 2{n^2}u} \right]} }}{{1 + {n^2}{u^2} \mp a\sqrt {{u^3}
\left[ {M\left( {1 - {n^2}{u^2}} \right) + 2{n^2}u} \right]} }}} \right)}}{{\left[ {\left( {{r^2} - 2Mr - {n^2}} \right) + 
{2a\left( {Mr + {n^2}} \right)}
\left( {\frac{{\sqrt {{u^3}\left[ {M\left( {1 - {n^2}{u^2}} \right) + 2{n^2}u} \right]} }}{{1 + {n^2}{u^2} \mp a\sqrt {{u^3}\left[ {M\left( {1 - {n^2}{u^2}} \right) 
+ 2{n^2}u} \right]} }}} \right)} \right]}}.
\end{equation}
These are shown in \figurename{5}.
By setting $a=0 $, we recover the  results for the NUT spacetime in the equatorial plane, for which the threading  and slicing 3-velocities  
are the same given by
\begin{equation}
v_S^\phi  = v_T^\phi  = \frac{{ \mp \sqrt {u\left[ {M\left( {1 - {n^2}{u^2}} \right) + 2{n^2}u} \right]} }}{{\sqrt {1 - 2Mu - {n^2}{u^2}}}}.
\end{equation}
%%%%%%%%%%%%%%%%%%%%%%%%%%%%%%%%
%%%%%%%%%%%%%%%%%%%%%%%%%%%%%%%%%%%%%%%%%%%
This is a direct consequence of the fact that the line element \eqref{kerr} for $a=0$ and in the equatorial plane becomes a static line element, and as pointed 
out earlier, in this 
case the two formalisms (and the two observers) coincide, leading to the same velocity definition. \\
For any other circular orbit ( i.e with $\theta \neq \frac{\pi}{2}$), even in the pure NUT case (when 
both mass and rotation parameters are zero), the two velocities would be different. Indeed 
the reason that we chose the Kerr-NUT spacetime as one of our examples is the fact that although in the case of Kerr metric one could attribute the relative 
linear velocity between the two sets of observers to the dragging
of frames (or the so called Lense-Thirring effect), that may not be so for other stationary spacetimes in which the cross terms are not necessarily originated 
from rotation. Indeed in the case of NUT metric, the NUT 
charge is mainly interpreted as representing a gravitomagnetic monopole,  the gravitational analogue of the Dirac monopole \cite{Misner,Lynden} and not a 
rotation parameter. So in general one could interpret this 
difference as a gravitomagnetic effect originated from the gravitomagnetic field of the underlying spacetime.
%%%%%%%%%%%%%%%%%%%%%%%%%%%%%%%%%%%%%%%%%%%%%%%%%%%%%%
\section{An astrophysical setup involving local velocity measurement}
\begin{figure}\label{6}
\includegraphics[scale=0.4]{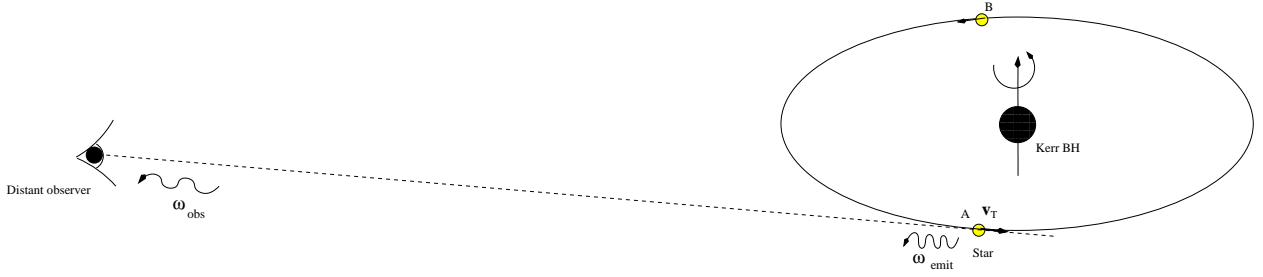}
\caption{A star orbiting a Kerr black hole in its equatorial plane in which also lies the line of sight of a distant observer receiving light from the star with 
frequency $\omega_{obs}$.}
\end{figure}

Having defined 3-velocity in curved backgrounds, it should be noted that these definitions of 3-velocity are given with respect to the
local observers in the sense that the particle and observer are close enough for the spacetime curvature to be neglected. In the lack of real observers sitting 
in the strong field zone, 
for distant observers these are basically theoretical definitions or as Synge puts it, 
``Mathematical Observations'' \cite{synge}. Hence one should try to relate them to the observational measurements of 3-velocity in astronomy 
and astrophysics which are made by distant observers at regions which may or may not be taken as asymptotically flat. \\
In astrophysics radial velocities are measured through the spectral line shifts of the light rays received from
stars but for their transverse velocities one should use their proper motion along with their distance.
Frequency shifts could be due to a combination of gravitational, cosmological as well as Doppler shifts. 
Here we picture an astrophysical scenario in which one could relate the emitted frequency in the rest frame of the star and that observed by a 
distant observer, through 
the frequency measured by the fundamental observer. 
Suppose there is a binary system either a double star or 
a black hole-star binary, with the  smaller star orbiting the much heavier rotating star/black hole around which the geometry is represented by the Kerr metric. 
An example could be stars orbiting the super massive black hole in the center of our own galaxy. For our purpose, 
and in  accordance with our calculations in section V-A, we take the star`s orbit 
to be a circle in the  Black hole's equatorial plane which in turn is looked at as an edge-on orbit by a distant observer (\figurename{6}).
%%%%%%%%%%%%%%%%%%%%%%%%%%%%%%%%%%%%
%%%%%%%%%%%%%%%%%%%%%%%%%
At different positions on the star's orbit the distant
observer receives different frequencies, specially for the light sent from the star at points on its orbit where
the observer's line of sight is tangent to the orbit, namely points A and B at which the star recedes and approaches radially both the 
fundamental and distant observers respectively (\figurename{6}). The relation between the observed and emitted frequencies 
could be derived in the following steps: \\
I-If the frequency of the light emitted by the orbiting star in its rest frame is denoted  $\omega_{emit}$ and that received by the 
fundamental observer is given by $\omega_{fo}$, then noting that for fundamental observers
the star's recessional velocity is equivalent to its Doppler velocity (as the curvature effects are taken to be negligible), we have
\begin{equation}
\frac{\omega_{emit}}{\omega_{fo}} =\sqrt{ \frac{1\pm v_T}{1\mp v_T}}
\end{equation}
where the upper and lower signs refer to the points A and B respectively.
%%%%%%%%%%%%%%%%%%%%%%%%%%%%
\begin{figure}\label{7}
\includegraphics[scale=1.2]{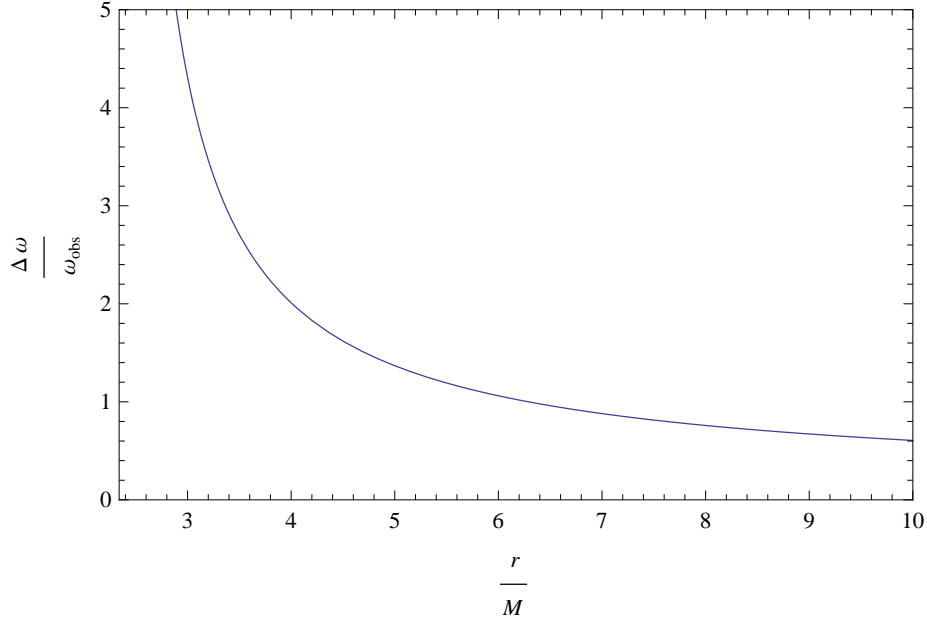}
\caption{Spectral shift measured by a distant observer in terms of the source's distance from the rotating black hole.}
\end{figure}
%%%%%%%%%%%%%%%%%%%%%%%%%%%%%%%%%%%%%%%%%%%
\\
II-If the frequency measured by the distant observer, residing at the asymptotically flat region, is denoted by  $\omega_{obs}$ and using the 
fact that the fundamental
observer and the star feel the 
same gravitational potential, the relation between the emitted and  observed frequency is given by,
\begin{equation}
{\omega^{A,B}_{obs}} = \sqrt{g_{00}(x^a_{star})} {\omega_{fo}} = \sqrt{g_{00}(x^a_{star})} \sqrt{ \frac{1 \mp v_T}{1 \pm v_T}} {\omega_{emit}}.
\end{equation}
Substituting for the $g_{00}$ from the Kerr metric in the equatorial plane we end up with
\begin{equation}
{\omega^{A,B}_{obs}} =\left(\sqrt{(1-2M/r)}\sqrt{ \frac{1\mp v_T (r)}{1\pm v_T (r)}}\right) {\omega_{emit}}.
\end{equation}
The spectral shift measured by a distant observer in terms of the source's normalized radial coordinate (at point A) is shown in \figurename{7}.
Since the gravitational redshift factor is the same everywhere on the circular orbit, the distant observer could use the frequency shift at the two 
points to eliminate  this factor and find that 
\begin{equation}
\frac {{\omega^A_{obs}}}{{\omega^B_{obs}}} =  \frac{1-v_T (r)}{1+v_T (r)}.
\end{equation}
%%%%%%%%%%%%%%%%%%%%%%%%%%%%%%%%
\section{Discussion}
In the present study, we considered two different definitions of 3-velocity of a test particle moving on timelike curves in a {\it stationary spacetime}. 
These two 3-velocities are defined in the context of the 
two well known spacetime decomposition formalisms, namely threading and  slicing formalisms. 
In terms of observers, they are defined relative to two different sets of observers, fundamental observers and Eulerian observers. 
The two sets of observers and their corresponding definitions of 3-velocities agree when the spacetime is static but differ when the spacetime 
is stationary having cross terms mixing space and time coordinates.
Restricting our attention to the axially symmetric stationary spacetimes which are naturally adapted to the astrophysical phenomena related mainly 
to rotating sources such as pulsars and Kerr black holes, we 
compare the two velocities as functions of radial coordinate of test particles in circular orbits around the corresponding spacetime  symmetry axis. 
By its definition 3-velocity is an observer-dependent quantity, fundamental observers in the threading approach are at rest with respect to a rigid 
global coordinate system, and hence are non-inertial 
observers. These observers use the proper time read by clocks synchronized along the particle's worldline and employ the corresponding spatial line 
element $dl_T$ to define the 3-velocity. 
In the case of Kerr 
metric these observers are called zero angular velocity observers (ZAVOs) and it is relative to these observers that the threading velocity of a 
test particle is defined.\\ 
In the ADM or slicing formalism, on the other hand, the 3-velocity is defined relative to the so called {\it fiducial observers} whose worldlines 
are orthogonal to the slicing hypersurfaces and so their 
definition depends on the chosen slicing. In the case of stationary axially symmetric spacetimes these observers are called {\it Eulerian observers}
in analogy with fluid mechanics since, although dragged 
by the geometry, they are  at rest on spatial slices. These observers are also non-inertial and due to the fact that their 4-velocity has a vanishing 
rotation (by their definition), are  
called {\it locally non-rotating observers} \cite{Bard}. In the case of Kerr metric, the same observers are called 
{\it zero-angular-momentum observers} (ZAMOs) \cite{Abra}.\\
It should be noted that although in the case of Kerr metric one could interpret the difference between the two velocities assigned to a particle 
in terms of the relative velocity between the two sets of observers
originating from the dragging of frames by the Kerr hole, that would not be the case in other stationary spacetimes such as the NUT spacetime whose 
cross terms are not traced back to a rotation parameter.
In the context of {\it gravitoelectromagnetism} one may assign this difference to the gravitomagnetic field of the underlying spacetime.\\
On the observational side one can think of a physical situation in which a  phenomena closely related to the rotational velocity of particles near 
massive rotating objects is being considered such as 
the phenomena related to the accretion disks around rotating black holes, in which matter swallowed by the black hole from its companion star forms
a rotating disk around its rotation axis. Another interesting 
example is the case of pulsars and related phenomena. The supermassive black hole in the center of our galaxy which harbors stars in near spherical 
orbits is another example in which the question of velocity 
measurement is an important issue.
In such cases due to the high rotational velocity of the object one should treat the underlying background as a Kerr spacetime and 
hence two different {\it local} velocities could be assigned to the rotating matter relative to two different sets of local observers. 
For example in the discussion on the relation between Penrose process \cite{Pen} and the energetics of superluminal jets 
emerging from quasars, the 3-velocity of disintegrating infalling particles are measured by ZAMO \cite{Bard,Wald}.
As shown in Fig. 4, for particles in orbits close to the marginally bound orbit, the difference between these local velocity measures 
and those made by the distant observers could be as high as $\%20$ percent of the velocity of light. Consequently, in considering the 
corresponding velocity-dependent phenomena  one should take into account which local velocity measure, if any, enters its formulation. 

%%%%%%%%%%%%%%%%%%%%%%%%%%%%%%%%%%%%%%%%%%%%%%%%%%%%%%
\section *{Acknowledgments}
R. Gharechahi  and M. Nouri-Zonoz thank University of Tehran for supporting this project under the grants provided by the research council. M.N-Z 
thanks the Albert Einstein Center for Fundamental 
Physics, University of Bern for  supporting his visit during which this study was carried out. M.N-Z also thanks Mathias Blau for useful discussions. 
Alireza Tavanfar thankfully 
acknowledges the support of his research by the Brazilian Ministry of Science, Technology and Innovation (MCTI-Brazil).
%%%%%%%%%%%%%%%%%%%%%%%%%%%%%%%%%%%%%%%%%%%%%%%%%%%%%%
%\pagebreak

%%%%%%%%%%%%%%%%%%%%%%%%%%%%%%%%%%%
%%%%%%%%%%%%%%%%%%%%%%%%%%%%%%%%%%%

\end{document}